\documentclass[12pt]{article}

\usepackage{epsfig,multirow,setspace}

\usepackage{setspace}
\usepackage{pdfpages}
\usepackage{rotating}

\usepackage{fancyhdr,url}
\usepackage{amsfonts, amsmath, amssymb, marvosym, upgreek, mathrsfs, dsfont} 
\usepackage{epstopdf}
\usepackage{array,graphicx,graphics}
\usepackage{booktabs}
\usepackage{pifont}
\usepackage{setspace}
\usepackage{colonequals} 
\usepackage{tikz,bm}
\usetikzlibrary{shapes,arrows,fit}
\usepackage[]{natbib}

\usepackage{enumitem}
\usepackage{caption}
\usepackage{subcaption}

\newcommand{\thetav}{\mbox{\boldmath{$\theta$}}}
\newcommand{\bY}{\mathbf{Y}}

\newcommand{\bPsi}{\boldsymbol{\Psi}}
\newcommand{\bOmega}{\boldsymbol{\Omega}}

\newcommand{\bDelta}{\boldsymbol{\Delta}}
\newcommand{\bKappa}{\boldsymbol{\kappa}}
\newcommand{\bxi}{\boldsymbol{\xi}}
\newcommand{\bPhi}{\boldsymbol{\Phi}}
\newcommand{\bs}{\mathbf{s}}
\def\vec{\text{vec}}
\newcommand{\M}{\mathbf{M}}
\newcommand{\by}{\mathbf{y}}
\DeclareMathOperator{\Ev}{\mathbb{E}}
\newcommand{\bI}{\mathbf{I}}
\newcommand{\diag}{\text{diag}}

\DeclareMathOperator{\E}{\mathbb{E}}
\DeclareMathOperator{\R}{\mathbb{R}}

\newcommand{\inv}{^{\raisebox{.2ex}{$\scriptscriptstyle-1$}}}

\usepackage{multirow}

\tikzstyle{decision} = [rectangle, draw, fill=red!20, 
    text width=10em, text centered, rounded corners,node distance=3cm, inner sep=2pt,minimum height=4em]
\tikzstyle{block} = [rectangle, draw, fill=blue!20, 
    text width=10em, text centered, rounded corners, minimum height=4em]
\tikzstyle{line} = [draw, -latex']
\tikzstyle{cloud} = [draw, ellipse,fill=red!20, node distance=3cm,
    minimum height=2em]

\begin{document}
	\doublespacing
	
	\title{\bf Finite Mixtures of Matrix Variate Poisson-Log Normal Distributions for Three-Way Count Data}
	
	\author{Anjali Silva\footnote{Department of Mathematics and Statistics, University of Guelph, Guelph, ON, Canada} \and Steven~J.~Rothstein\footnote{Department of Molecular and Cellular Biology, University of Guelph, Guelph, ON, Canada} \and Paul~D.~McNicholas\footnote{Department of Mathematics and Statistics, McMaster University, Hamilton, ON, Canada} \and Xiaoke Qin\footnote{School of Mathematics and Statistics, Carleton University, Ottawa, ON, Canada} \and Sanjeena Subedi \footnote{School of Mathematics and Statistics, Carleton University, Ottawa, ON, Canada. e: sanjeena.dang@carleton.ca}}
	\date{}

	\maketitle

\begin{abstract}
Three-way data structures, characterized by three entities, the units, the variables and the occasions, are frequent in biological studies. In RNA sequencing, three-way data structures are obtained when high-throughput transcriptome sequencing data are collected for $n$ genes across $p$ conditions at $r$ occasions. Matrix variate distributions offer a natural way to model three-way data and mixtures of matrix variate distributions can be used to cluster three-way data. Clustering of gene expression data is carried out as means of discovering gene co-expression networks. In this work, a mixture of matrix variate Poisson-log normal distributions is proposed for clustering read counts from RNA sequencing. By considering the matrix variate structure, full information on the conditions and occasions of the RNA sequencing dataset is simultaneously considered, and the number of covariance parameters to be estimated is reduced. We propose three different frameworks for parameter estimation: a Markov chain Monte Carlo based approach, a variational Gaussian approximation based approach, and a hybrid approach. Various information criteria are used for model selection. The models are applied to both real and simulated data, and we demonstrate that the proposed approaches can recover the underlying cluster structure in both cases. In simulation studies where the true model parameters are known, our proposed approach shows good parameter recovery.
\end{abstract}

% keywords can be removed
%\textbf{Keywords}:Model-based clustering, Biclustering, AECM, Factor analysis, Mixture models
%
%\footnotetext{\textbf{Abbreviations:} AECM, alternating expectation conditional maximization; ALL, acute lymphoblastic leukemia; AML, acute myeloid leukemia; ARI, adjusted Rand index; BIC, Bayesian information criteria; EM, expectation-maximization; MFA, mixtures of factor analyzers; FDR, false discovery rate}

\section{Introduction}

\label{sec:intro}
Finite mixture models are popular for clustering applications and are widely used on two-way data \citep{mclachlan2000, mcnicholas2016}. Three-way data are becoming increasingly commonplace in several fields, including bioinformatics. Three-way data structures are characterized by three entities or modes: the units (rows), the variables (columns), and the occasions (layers). For two-way data, each observation is represented as a vector whereas, for three-way data, each observation can be regarded as a matrix. A random matrix $\mathbf{T}_n$ is said to contain $k \in \{1,\ldots,p\}$ responses over $i \in \{1,\ldots,r\}$ occasions and $n = 1, \ldots, N$ such units are considered. This provides $N$ independent and identically distributed random matrices $\mathbf{T}_1, \mathbf{T}_2, \ldots, \mathbf{T}_N$.

Matrix variate distributions offer a natural approach for modeling three-way data. 
Extensions of matrix variate distributions in the context of mixture models have given rise to mixtures of matrix variate distributions, which have been used to cluster three-way data \citep{Viroli2011, Anderlucci2015, Dogru2016, Gallaugher2017,Gallaugher2018}. Here, the interest lies in clustering the $N$ observed matrices into $G$ clusters, while utilizing all information from the other two modes \citep{Viroli2011}. It is assumed that matrices $\mathbf{T}_1, \mathbf{T}_2, \ldots, \mathbf{T}_N$ are conditionally independent and identically distributed observations coming from a mixture model with $G$ possible groups in proportions $\pi_{1},\ldots ,\pi_{G}$ \citep{Viroli2011}. The density of the $G$-component mixture is $ f( \mathbf{T}_N|\pi_{1}, \ldots,\pi_{G},\boldsymbol{\vartheta}_1,\ldots, \boldsymbol{\vartheta}_G ) = \sum_{g=1}^{G} \pi_{g} f^{(r \times p)}( \mathbf{T}_n | \boldsymbol{\vartheta}_g).$ Here parameters of the distribution function $f^{(r \times p)}(\cdot)$ are represented by $\boldsymbol{\vartheta}_g$ and $\pi_{g}>0$, such that $\sum_{g=1}^G \pi_g=1$, is the mixing proportion of the $g$th component.

Three-way datasets are common in biological studies, including RNA sequencing (RNA-seq), where gene expression count data are collected for $N$ genes across $p$ conditions at $r$ occasions. \citet{Silva2019} proposed a mixture model-based clustering methodology based on the multivariate Poisson-log normal (MPLN) distribution for two-way RNA-seq data. For genes $n \in \{1, \ldots, N\}$ and samples $c \in \{1, \ldots, rp\}$, the MPLN distribution is given by
\begin{equation*}\begin{split}
\label{eqn:MPLN2}
Y_{nc} | \theta_{nc} &\sim \mathscr{P}(\exp\{\theta_{nc}+{\log s_{c}}\} ) \\
(\theta_{n1},\ldots,\theta_{nrp})^{\prime} &\sim \mathcal{N}_{rp}(\boldsymbol{\mu},\mathbf{\Sigma}),
\end{split}
\end{equation*}
where $\mathscr{P}$ denotes the Poisson distribution and the $\mathcal{N}_{rp} $ is a $rp$-dimensional normal distribution. To account for the differences in library sizes across each sample $c$ of an RNA-seq study, a fixed, known constant $s_c$, representing the normalized library sizes, is added to the mean of the Poisson distribution. In this work, mixtures of MPLN distributions and matrix variate normal distributions are extended to give rise to mixtures of matrix variate Poisson-log normal (MVPLN) distributions for clustering three-way count data. Details of parameter estimation are provided, and both real and simulated data illustrations are used to demonstrate the clustering ability.

\section{Methods}
\label{sec:meth}
\subsection{Matrix variate Poisson-log normal distribution}
Mathematical properties of the matrix variate normal distribution can be found in \citet{gupta2000}. By considering a matrix variate structure, the number of free covariance parameters to be estimated is reduced from $\frac{1}{2} rp(rp + 1)$ to $ \frac{1}{2} [r(r + 1) + p(p + 1)]$. The matrix variate normal distribution can be extended to give rise to MVPLN distribution using a hierarchical structure. Consider $n$ independent and identically distributed random matrices $\mathbf{Y}_n$, $n = 1, \ldots, N$, each of dimension $r \times p$. In the MVPLN frameword, $Y_{nik}| \theta_{nik}$ follows a Poisson distribution with mean $\text{exp}(\theta_{nik})$, and $(\boldsymbol{\theta}_{n})^{\prime}$ follows a $r \times p$ matrix variate normal distribution $\mathcal{N}_{r \times p} (\mathbf{M}, \mathbf{\Phi}, \mathbf{\Omega})$, where $\mathbf{M}$ is a $r \times p$ matrix of means, $\mathbf{\Phi}$ is a $r \times r$ covariance matrix containing the variances and covariances between $r$ occasions and $\mathbf{\Omega}$ is a $p \times p$ covariance matrix containing the variance and covariances of the $p$ variables. Figure \ref{graph} provides a graphical representation of a mixture of MVPLN distributions.
\begin{figure}[!th]
 \scalebox{.8}{
\begin{tikzpicture}[node distance = 4cm, auto]
%Placing nodes
  \node [decision] (mod) {$Z_{ng}$\\ (Latent variable)};
\node [cloud, left of=mod, node distance=5cm] (pi) {$\pi_g$};
    \node [decision, below of=mod,node distance=2.5cm] (theta) { $\bm{\theta}_{n}$ \\(Latent variable)};
\node [cloud, below of=pi,node distance=1.2cm] (mu) {$\bm{M}_{g}$};
    \node [cloud, below of=mu,node distance=1.2cm] (phi) {$\bm{\Phi}_g$};
      \node [cloud, below of=phi,node distance=1.2cm] (omega) {$\bm{\Omega}_g$};
  
          \node [block, right of=mod,node distance = 6cm,right of=mu] (Y) {$\mathbf{Y}_{n}$ \\(Observed data)};
        % Drawing edges
         \path [line] (pi) -- (mod);
          \path [line] (mu) -- (theta);
           \path [line] (phi) -- (theta);
            \path [line] (omega) -- (theta);
            \path [line] (mod) -- (theta);
              \path [line] (mod) -- (Y);
              \path [line] (theta) -- (Y);      
              
              \draw (-4.5,-1.85)  node[minimum height = 2.25in, minimum width=1 in,draw] {\parbox[b][2.25in]{1in}{$g=1,\ldots,G$}}; 
                  \draw (3.5,-1.5)  node[minimum height = 2in, minimum width=5in,draw] {\hspace{3in}\parbox[b][1.5in]{1in}{$n=1,\ldots,N$}};
\end{tikzpicture}}
\caption{Graphical representation of the MVPLN mixture model.}\label{graph}
\end{figure}
 
The vectorization of $\mathbf{Y}_n$, denoted $\text{vec}(\mathbf{Y}_n)$, is $rp$-dimensional.Given all $\text{vec}(\mathbf{Y}_n)$, i.e., for $n=1, \ldots, N$, the library sizes $\text{vec}(\mathbf{s})$ can be calculated. The $\text{vec}(\mathbf{s})$ and $\text{vec}(\boldsymbol{\theta}_{n})$ are $rp$-dimensional. The covariance matrix of $\text{vec}(\mathbf{Y}_n)$ is $\mathbf{\Sigma} = \mathbf{\Phi} \otimes \mathbf{\Omega}$, where $\otimes$ denotes the Kronecker product. Note that $\mathbf{\Sigma}$ has dimension $rp \times rp$, and the probability mass function of the MVPLN distribution is
\begin{equation*}\begin{split}
&f(\mathbf{Y}_n, \mathbf{s} | \boldsymbol{\vartheta}) = \int_{\R} \left\{ \prod _{c=1}^{rp} f(\text{vec}(\boldsymbol{Y}_{n})_{c} | \text{vec}(\boldsymbol{\theta}_{n})_{c}, \text{vec}(\mathbf{s})_c) \right\} g^{(r \times p)} (\boldsymbol{\theta}_{n} | \boldsymbol{\vartheta})~d\text{vec}(\boldsymbol{\theta}_{n}),
\end{split}
\end{equation*}
where $\boldsymbol{\vartheta} = (\mathbf{M}, \mathbf{\Phi}, \mathbf{\Omega})$, $f(\cdot)$ is the probability mass function of Poisson distribution and $g^{(r \times p)}(\cdot)$ is the probability density function of matrix variate normal distribution.

The unconditional mean and covariance of the MPLN distribution can be calculated using the properties of the log-normal distribution and of the conditional expectation \citep{aitchison1989, tunaru2002}. For the MVPLN distribution, the unconditional mean and covariance are
\begin{equation*}\begin{split}
 \E(Y_{ik}) &= \E [\E(Y_{ik}|\theta_{ik})] = \exp\left\{\mu_{ik} + \frac{1}{2}(\mathbf{\Phi}_{ii} \mathbf{\Omega}_{kk})\right\} = \mathcal{M}_{ik},\\
 \mathbb{V}\text{ar}(Y_{ik}) & =  \E [\mathbb{V}\text{ar}(Y_{ik}|\theta_{ik})] + \mathbb{V}\text{ar} [\E(Y_{ik}|\theta_{ik})] \\
 & = \mathcal{M}_{ik} + \mathcal{M}_{ik}^{2}\left(\exp\{\mathbf{\Phi}_{ii} \mathbf{\Omega}_{kk}\}-1\right),
\end{split}
\end{equation*} respectively.
The MVPLN distribution can account for both the correlations between variables and the correlations between occasions, as two different covariance matrices are used for the two modes. This makes the model ideal for modeling RNA-seq data when expression measurements for different conditions at different occasions, e.g., time-points or replicates, are available.

\subsection{Finite mixtures of MVPLN distributions}
\begin{figure}[!h]%figure1
\centerline{\includegraphics[width=0.5\textwidth]{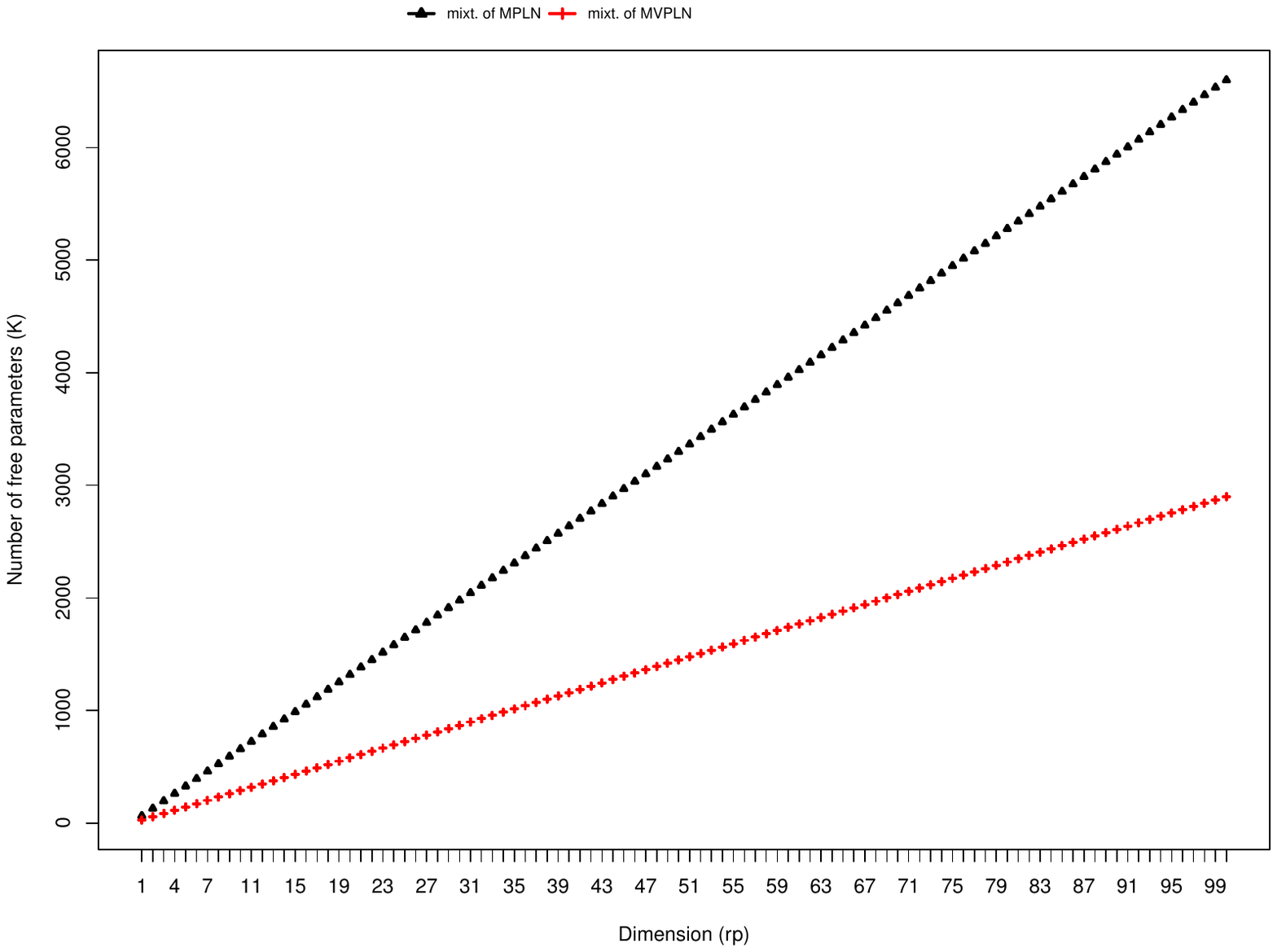}}
\caption{Scatter plot illustrating how the number of free parameters $K$ grows with data dimensionality $rp$ for the mixtures of MPLN model and for the mixtures of MVPLN model. Here $G = 2, r =2,$ and $rp = 4 \text{ up to } 100$.} \label{freepara1}
\centerline{\includegraphics[width=0.5\textwidth]{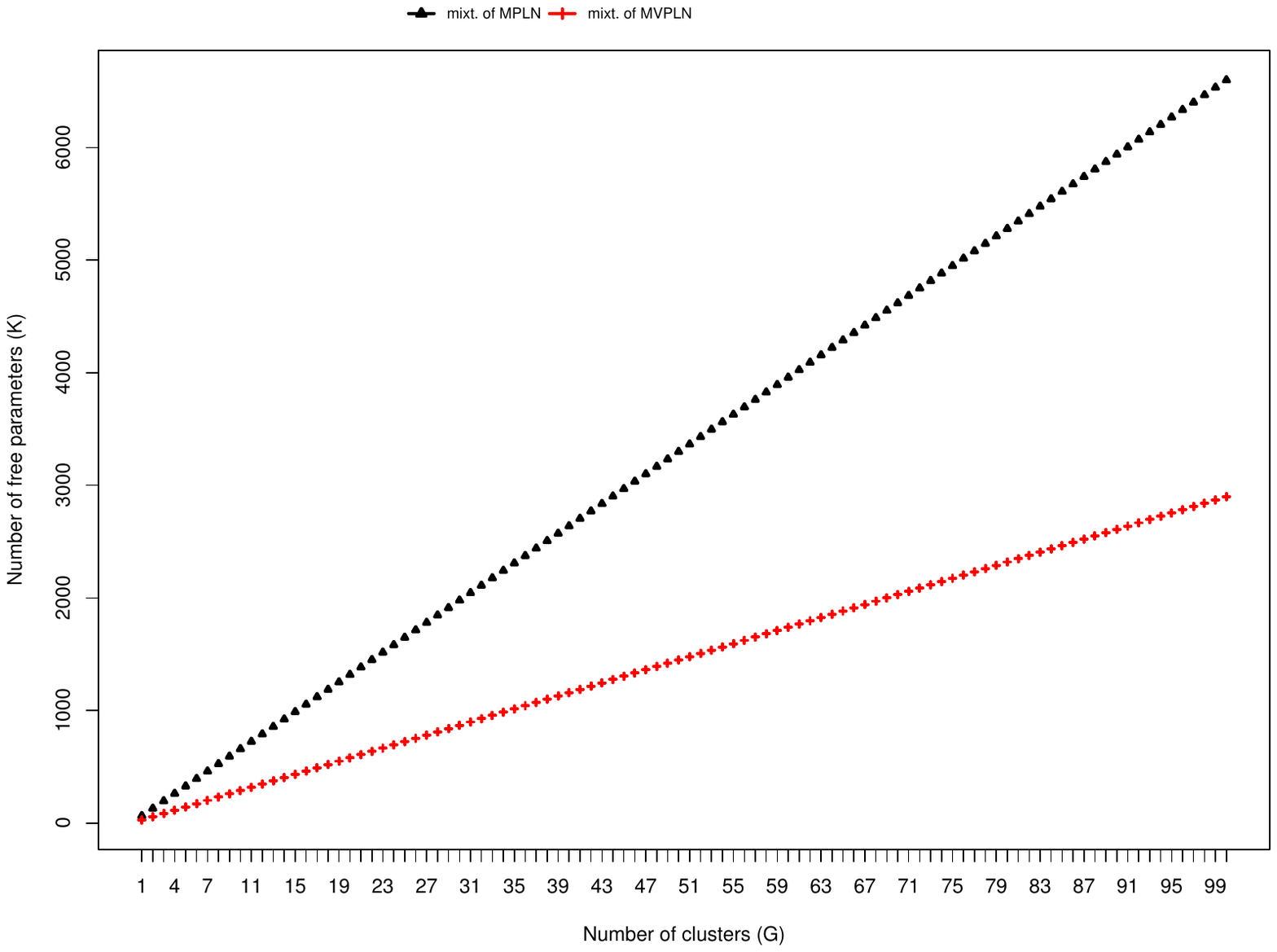}}
\caption{Scatter plot illustrating how the number of free parameters $K$ grows with the number of clusters $G$ for the mixtures of MPLN model and for the mixtures of MVPLN model. Here $G = 1:100, r = 2, p = 5$.} \label{freepara2}
\end{figure}

In the mixture model context, a random matrix $\mathbf{Y}_n$ is assumed to come from a population with $G$ subgroups each distributed according to an MVPLN distribution. Then $N$ such matrices $\mathbf{Y}_1, \mathbf{Y}_2,\ldots,\mathbf{Y}_N$ are observed, each of which belongs to one of $g \in \{1,\ldots, G\} $ different sub-populations with mixing proportions $\pi_1,\ldots,\pi_G$. Then the probability density function of a $G$-component mixture of MVPLN distributions can be written as
\vspace{-0.1in}
\begin{equation*}
\begin{split}
f(\mathbf{Y};\boldsymbol{\Theta}) &= \sum_{g=1}^G\pi_gf_{\mathbf{Y}}(\mathbf{Y}| \mathbf{M}_g,\mathbf{\Phi}_g, \mathbf{\Omega}_g) \\ &=  \sum_{g=1}^{G} \pi_g \int_{\R} \left\{\prod_{c=1}^{rp} f_g \Big( \text{vec}(\mathbf{Y}_n)_{c} | \text{vec}(\boldsymbol{\theta}_{ng})_{c}, \text{vec}(\mathbf{s})_c \Big) \right\} \\
& \quad\qquad\qquad\qquad\qquad\times g_g^{(r \times p)} (\boldsymbol{\theta}_{ng} | \mathbf{M}_g,\mathbf{\Phi}_g, \mathbf{\Omega}_g)~d \text{vec}(\boldsymbol{\theta}_{ng}),
\end{split}
\end{equation*}
where $\boldsymbol{\Theta} = (\pi_1, \ldots, \pi_G, \mathbf{M}_1, \ldots, \mathbf{M}_G, \mathbf{\Phi}_1, \ldots, \mathbf{\Phi}_G, \mathbf{\Omega}_1, \ldots, \mathbf{\Omega}_G)$, the $f_g(\cdot)$ is the probability mass function of a Poisson distribution and the $g_g^{(r \times p)}(\cdot)$ is the probability density function of matrix variate normal distribution. The cluster membership of all units is assumed to be unknown and $z_{ng}$ is used to cluster membership, where $z_{ng}=1$ if $\mathbf{Y}_n$ is in component $g$ and $z_{ng}=0$ otherwise. The complete-data consist of the observed and missing data, i.e., ($\mathbf{Y}_1,\ldots,\mathbf{Y}_N, \boldsymbol{z}_1,\ldots,\boldsymbol{z}_N,\boldsymbol{\theta}_1,\ldots,\boldsymbol{\theta}_N$). The complete-data likelihood is
\begin{equation*} \begin{split}
L_c (\boldsymbol{\Theta}) = \prod_{n=1}^N \prod_{g=1}^G \Bigg[ \pi_g \left\{ \prod _{c=1}^{rp} f_g \Big( \text{vec}(\mathbf{Y}_n)_{c} | \text{vec}(\boldsymbol{\theta}_{ng})_{c}, \text{vec}(\mathbf{s})_c \Big) \right\} \\
\times g_g^{(r \times p)} (\boldsymbol{\theta}_{ng} | \mathbf{M}_g,\mathbf{\Phi}_g, \mathbf{\Omega}_g) \Bigg]^{z_{ng}},
\end{split}
\end{equation*}
and the complete-data log-likelihood is
\begin{equation*}
\begin{split}
l_c& (\boldsymbol{\Theta})  = \sum_{g=1}^G n_g \log \pi_g\\
& - \sum_{n=1}^N \sum_{g=1}^G  \sum_{c=1}^{rp} z_{ng} \exp\{\text{vec}(\boldsymbol{\theta}_{ng})_{c} + \log \text{vec}(\mathbf{s})_c\} \\
& + \sum_{n=1}^N \sum_{g=1}^G z_{ng} \left[\text{vec}(\boldsymbol{\theta}_{ng}) + \log \text{vec}(\mathbf{s})\right]\text{vec}(\mathbf{Y}_n)^{\prime} \\
& - \sum_{n=1}^N \left(\sum_{g=1}^G z_{ng} \right) \sum_{c=1}^{rp} \log (\text{vec}(\mathbf{Y}_n)_{c})! - \frac{nrp}{2} \log (2\pi) \\
& - \frac{p}{2} \sum\limits_{g=1}^G n_g \log |\mathbf{\Phi}_g| - \frac{r}{2} \sum\limits_{g=1}^G n_g \log |\mathbf{\Omega}_g| \\
& - \frac{1}{2} \sum\limits_{n=1}^N \sum\limits_{g=1}^G z_{ng} \text{tr} \left[ \mathbf{\Phi}_g^{\inv} (\boldsymbol{\theta}_{ng} - \mathbf{M}_g)\mathbf{\Omega}_g^{\inv}(\boldsymbol{\theta}_{ng} - \mathbf{M}_g)^{\prime} \right],
\end{split}
\end{equation*}
where $n_g = \sum_{n=1}^N z_{ng}$. Compared to the mixtures of MPLN distribution, the number of free parameters to be estimated is reduced by considering a matrix variate structure (see Figures \ref{freepara1} and \ref{freepara2}). For the mixtures of MPLN model, the number of free parameters is $K = (G-1) + (Grp) + \frac{1}{2}Grp [rp +1] $, whereas for mixtures of MVPLN model it is $K = (G-1) + (Grp) + \frac{1}{2}G [ r(r+1) + p(p+1)]$.

\subsection{Parameter estimation}\label{sec:para}
Three different frameworks for parameter estimation for the mixtures of MVPLN models are proposed: one based on Markov chain Monte Carlo (MCMC) methods, one based on variational Gaussian approximations (VGAs) as well as a hybrid approach. MCMC-based approaches are computationally intensive; hence, we also provide a computationally efficient variational approximation framework for parameter estimation. Finally, a hybrid approach combines the variational approximation based approach and MCMC based approach.
\subsubsection{MCMC based approach}
In the MCMC based approach, the Markov chain Monte Carlo expectation-maximization (MCMC-EM) algorithm is used to estimate the model parameters \citep[see][for details]{Silva2019}. Using an MCMC-EM algorithm, the expected value of the $\boldsymbol{\theta}_{ng}$ and the $Z_{ng}$ conditional on the parameter updates from the $t^{th}$ iteration, respectively, are updated in the expectation (E-) step  as follows:
\begin{equation}
\begin{split}
\label{eqn:Ezig_chapter5}
& \E ( \boldsymbol{\theta}_{ng} | \mathbf{Y}_n) \simeq \frac{1}{W} \sum_{f=1}^W \boldsymbol{\theta}_{ng}^{(f)} \simeq \boldsymbol{\theta}_{ng}^{(t)}, \\
& \E (Z_{ng} |\mathbf{Y}_n,  \boldsymbol{\theta}_{ng}, \mathbf{s} ) = \frac{q_{ng}}{\sum_{h=1}^G q_{nh}} := z_{ng}^{(t)},
\end{split}
\end{equation}
where 
\begin{equation*}
\begin{split}
q_{ng}= \pi_g^{(t)}&\left\{\prod _{c=1}^{rp} f_g \left(\text{vec}(\mathbf{Y}_n)_{c} | \text{vec}(\boldsymbol{\theta}_{ng})_{c}, \text{vec}(\mathbf{s})_c \right)\right\}g_g^{(r \times p)} (\boldsymbol{\theta}_{ng}^{(t)} | \boldsymbol{M}_g^{(t)}, \boldsymbol{\Phi}_g^{(t)}, \boldsymbol{\Omega}_g^{(t)})
\end{split}
\end{equation*}
and $ \boldsymbol{\theta}_{ng}^{(f)}$ is a random sample simulated via the ${\tt RStan}$ package for iterations $f = 1,\ldots, B$. In the E-step, the expectation is taken conditional on the current parameter estimates; hence, the use of $(t)$ on parameters in \eqref{eqn:Ezig_chapter5}. As the values from initial iterations are discarded from further analysis to minimize bias, the number of iterations used for parameter estimation is $W$, where $W < B$.  The conditional expected-value of the complete-data log-likelihood is
{\small\begin{equation*}
\begin{split}
& \mathcal{Q}(\boldsymbol{\Theta})  \simeq \E \Big\{l_c (\boldsymbol{\Theta} ) \Big\}
 \simeq C - \frac{p}{2} \sum\limits_{g=1}^G n_g \log |\mathbf{\Phi}_g| - \frac{r}{2} \sum\limits_{g=1}^G n_g \log |\mathbf{\Omega}_g| \\
& -\! \frac{1}{2}\! \sum\limits_{n=1}^N\! \sum\limits_{g=1}^G\! z_{ng}^{(t)}\! \E\!  \Big[ \text{tr} \Big(\mathbf{\Phi}_g^{\inv}\! (\boldsymbol{\theta}_{ng}\! - \mathbf{M}_g\!) \mathbf{\Omega}_g^{\inv}\!(\boldsymbol{\theta}_{ng}\! -\! \mathbf{M}_g\!)^{\prime}\!\Big) | \boldsymbol{\theta}^{(t)}_{ng},\! z_{ng}\! =\! 1\! \Big],
\end{split}
\end{equation*}}where $C$ is a constant with respect to $\boldsymbol{M}_g$, $\boldsymbol{\Phi}_g$ and $\boldsymbol{\Omega}_g$, and $n_g^{(t)} = \sum_{n=1}^N z_{ng}^{(t)}$. During the M-step, the updates for the parameters are obtained as follows:
{\small\begin{equation*}
\label{eqn:Mpi}
\small \pi_g^{(t+1)} = \frac{n_g^{(t)}}{N}, \qquad
\small \boldsymbol{M}^{(t+1)}_{g} = \frac{\sum_{n=1}^N z_{ng}^{(t)} \E\big(\boldsymbol{\theta}_{ng}\big)} {n_g},
\end{equation*}
\begin{equation*}
\label{eqn:Mphi}
\small \mathbf{\Phi}^{(t+1)}_g = \frac{\sum_{n=1}^N z_{ng}^{(t)} \E\Big( (\boldsymbol{\theta}_{ng} - \mathbf{M}^{(t+1)}_g) (\mathbf{\Omega}_g^{(t)})^{\inv} (\boldsymbol{\theta}_{ng} - \mathbf{M}^{(t+1)}_g)^{\prime}\Big)}{p n_g},
\end{equation*}
\begin{equation*}
\label{eqn:Momega}
\small \mathbf{\Omega}^{(t+1)}_{g} = \frac{\sum_{n=1}^N z_{ng}^{(t)} \E\Big ( (\boldsymbol{\theta}_{ng} - \mathbf{M}^{(t+1)}_g)^{\prime} (\mathbf{\Phi}_g^{(t+1)})^{\inv} (\boldsymbol{\theta}_{ng} - \mathbf{M}^{(t+1)}_g)\Big) }{r n_g}.
\end{equation*}}

\subsubsection{VGA based approach}

Variational approximations \citep{wainwright2008} are approximate inference techniques in which a computationally convenient approximating density is used in place of a more complex but `true' posterior density. The approximating density is obtained by minimizing the Kullback-Leibler (KL) divergence between the true and the approximating densities. Suppose we have an approximating density $q(\boldsymbol{\theta})$, then the marginal log of the probability mass function can be written
\begin{align*}
\log f_Y(\mathbf{Y}) &= F(q,\mathbf{Y}) + D_{KL}(q\|f_Y),
\end{align*}
\noindent where $D_{KL}(q\|f)= \int q(\thetav) \log \frac{q(\boldsymbol{\theta})}{f(\thetav\mid\mathbf{Y})} d\boldsymbol{\theta}$ is the KL divergence between $f(\thetav\mid \bY)$ and approximating distribution $q(\thetav)$, and $F(\bY, q)=\int [\log f(\bY, \thetav)-\log q(\thetav)] q(\thetav) d \thetav$ is our evidence lower bound (ELBO). 
Thus, to minimize the KL divergence, we maximize our ELBO. For variational Gaussian approximations, $q(\thetav)$ is assumed to be a Gaussian distribution.

The complete-data log-likelihood for the mixtures of MVPLN distributions can be written
\begin{align*}
l_c(\boldsymbol{\Theta}) &=\sum_{g=1}^G \sum_{n=1}^N z_{ng} \log \pi_g + \sum_{g=1}^G \sum_{n=1}^N z_{ng} \log f(\bY_n\mid \M_g,\bPhi_g,\bOmega_g) \\
= \sum_{g=1}^G& \sum_{n=1}^N z_{ng} \log \pi_g +  \sum_{g=1}^G \sum_{n=1}^N z_{ng} \left[ F(\bY_n,q_{ng}) + D_{KL}(q_{ng}\|f_{ng}) \right],
\end{align*}
 \noindent where $D_{KL}(q_{ng}\|f_{ng})= \int q(\thetav_{ng}) \log \frac{q(\thetav_{ng})}{f(\thetav_{n}\mid\bY_n,Z_{ng}=1)} d\thetav_{ng}$ is the KL divergence between $f(\thetav_n\mid \bY_n,Z_{ng}=1)$ and approximating distribution $q(\thetav_{ng})$. Assuming $q(\thetav_{ng})=N_{r\times p}(\bxi_{ng},\bDelta_{ng},\bKappa_{ng})$, the ELBO for each observation $\by_n$ becomes
\begin{align*}
& F(q_{ng},\bY_n) = - \left[ \sum_{i=1}^{r}\sum_{j=1}^{p}\exp \left\{   (\bxi_{ng})_{ij}+\frac{1}{2}(\bDelta_{ng})_{ii}(\bOmega_{ng})_{jj}\right.\right.\\
& +\log\bs_{ij}  \left.\bigg\} \right]+ \left[\text{vec}\left(\bxi_{ng}^\top\right)+\log\vec(\bs)\right]^\top \vec(\bY_n^\top) \\
&  - \left[ \sum_{c=1}^{rp}\log(\vec(\bY_n^\top)_c!)\right] -\frac{p}{2}  \log |\bPhi_g|  -\frac{r}{2}  \log |\bOmega_g|\\
 &-\frac{1}{2}\left( \vec(\bxi_{ng}^\top)-\vec(\M_g^\top)\right)^\top\bPhi_g^{-1}\otimes\bOmega_g^{-1}\left(\vec(\bxi_{ng}^\top)-\vec(\M_g^\top)\right)\\
&+ \text{tr}\left\{\bPhi_g^{-1} \bDelta_{ng}\right\}\text{tr}\left\{\bOmega_g^{-1}\bKappa_{ng}\right\} + \frac{p}{2}  \log |\bDelta_{ng}| + \frac{r}{2}  \log |\bKappa_{ng}| + \frac{r p}{2}.
 \end{align*}
 
 The variational parameters that maximize the ELBO will minimize the KL divergence between the true posterior and the approximating density. Thus, parameter estimation can be done in an iterative EM-type approach such that the following steps are iterated. At the $(t+1)^{th}$ step, 
\begin{enumerate}[noitemsep,leftmargin=*]
\item Conditional on the variational parameters $\bxi_{ng}$, $\bDelta_{ng}$, and $\bKappa_{ng}$  and on $\M_g$, $\bPhi_g$ and $\bOmega_g$, the  $\Ev(Z_{ng})$ is computed. Given $\pi_g$, $\M_g$, $\bPhi_g$ and $\bOmega_g$, 
\begin{align*}
\Ev(Z_{ng} \mid \bY_n)=\frac{\pi_g f(\bY\mid \M_g, \bPhi_g ,\bOmega_g)}{\sum_{h=1}^G\pi_h f(\bY_n\mid \M_h, \bPhi_h ,\bOmega_h)}.
\end{align*}
Note that this involves the marginal distribution of $\bY$ which is difficult to compute. Hence, we use an approximation of $\Ev(Z_{ng})$, where we replace the marginal density of the exponent of ELBO such that 
\begin{align*}
\widehat{Z}_{ng}^{(t+1)}\stackrel{\text{def}}{=}\frac{\pi_g \exp\left[F\left(q_{ng},\bY_n\right) \right]}{\sum_{h=1}^G\pi_h \exp\left[F\left(q_{nh},\bY_n\right) \right]}.
\end{align*}
This approximation is computationally convenient and a similar framework has been previously utilized \citep{gollini2014,tang2015}. This approximation works well in simulation studies and real data analysis.

\item \ Given $\hat{Z}_{ng}^{(t+1)}$, variational parameters  $\bxi_{ng}$, $\bDelta_{ng}$, and $\bKappa_{ng}$ are updated conditional on $\M_g^{(t)}$, $\bPhi_g^{(t)}$ and $\bOmega_g^{(t)}$.
\begin{enumerate}[leftmargin=*]
 \item \ A fixed-point method is used for updating $\bDelta_{n}$:
 \begin{align*}
&\bDelta_{ng}^{(t+1)}=p\Bigg[\bI_{r\times r} \odot \bigg\{\diag(\bKappa_{ng}^{(t)})^\top \bigg[\mathbf{exp}\bigg\{\bxi_{ng}^{(t)}+\log \bs \\
&\left.\left.\left.\left.+\frac{1}{2}\diag(\bDelta_{ng}^{(t)})\diag(\bKappa_{ng}^{(t)})^\top\right\}\right]^\top\right\}+\bPhi_g^{-1(t)}\text{tr}\left\{\bOmega_g^{-1(t)}\bKappa_{ng}^{(t)}\right\} \right]^{-1},
\end{align*}

where the vector function $\mathbf{exp}\left[ \mathbf{a} \right] = (e^{a_1}, \ldots, e^{a_{r}})'$ is a vector of the exponential of each element of the $r$-dimensional vector $\mathbf{a}$, $\mbox{diag}( \bKappa ) = ( \bKappa_{11} \ldots,  \bKappa_{pp})$ puts the diagonal elements of the $p\times p$ matrix  $\bKappa$ into a $p$-dimensional vector, and $\odot$ is the Hadmard product.

 \item A fixed-point method is used for updating $\bKappa_{ng}$:
 \begin{align*}
&\bKappa_{ng}^{(t+1)}=r\left[  \bI_{p \times p} \odot \left\{ \diag(\bDelta_{ng}^{(t+1)})^\top \left[\mathbf{exp}\bigg\{\bxi_{ng}^{(t)}+\log \bs \right.\right.\right.\\
&\left.\left.\left.\left.+\frac{1}{2}\left(\diag(\bKappa_{ng}^{(t)})\diag(\bDelta_{ng}^{(t+1)})^\top\right)^\top\right\}\right]\right\}+\bOmega_g^{-1(t)}\text{tr}\left\{\bPhi_g^{-1(t)}\bDelta_{ng}^{(t+1)}\right\} \right]^{-1},
\end{align*}
where the vector function $\mathbf{exp}\left[ \mathbf{a} \right] = (e^{a_1}, \ldots, e^{a_{p}})'$ is a vector of exponential each element of the $p$-dimensional vector $\mathbf{a}$, $\mbox{diag}( \bDelta ) = ( \bDelta_{11} \ldots,  \bDelta_{rr})$ puts the diagonal elements of the $r\times r$ matrix  $\bDelta$ into a $r$-dimensional vector, and $\odot$ is the Hadmard product.

\item \ Newton's method is used to update $\bxi_{ng}$:
\begin{align*}
 \vec(\bxi_{ng}^{\top(t+1)})&=\vec(\bxi_{ng}^{\top(t)}) -\bPsi_{ng}^{-1(t+1)} \left\{ \vec(\bY^\top_n) - \mathbf{exp}\left[\log \vec(\bs^\top)\right.\right.\\
&+\vec(\bxi_{ng}^{\top(t)}) \left.+\frac{1}{2}\mbox{diag}\left(\bPsi_{ng}^{-1 (t+1)}\right)\right]\\
&- \bPsi_{ng}^{-1(t+1)}\left(\vec(\bxi_{ng}^{\top(t)})-\vec(\M^{\top(t)}_g)\right)\left.-\vec(\bY^\top_n)\right\},
\end{align*}
where $\bPsi_{ng}^{(t+1)}=\bDelta_{ng}^{(t+1)}\otimes\bKappa_{ng}^{(t+1)}$.
\end{enumerate}

\item \ Given $\hat{Z}_{ng}^{(t+1)}$ and the variational parameters $\bxi_{ng}^{(t+1)}$, $\bDelta_{ng}^{(t+1)}$, and $\bKappa_{ng}^{(t+1)}$ , the parameters $\pi_g$, $\M_g$, $\bPhi_g$ and $\bOmega_g$ can be solved for as
\begin{align*}
\pi_g^{(t+1)} &= \frac{n_g^{(t+1)}}{N}~\text{where}~ n_g^{(t+1)}= \sum_{n=1}^N\hat{Z}_{ng}^{(t+1)}, \\
\boldsymbol{M}^{(t+1)}_{g} &= \frac{\sum_{n=1}^N \hat{Z}_{ng}^{(t+1)} \bxi_{ng}^{(t+1)}} {n_g^{(t+1)}},\\
\mathbf{\Phi}^{(t+1)}_g &= \frac{\sum_{n=1}^N \hat{Z}_{ng}^{(t+1)} (\bxi_{ng}^{(t+1)} - \mathbf{M}^{(t+1)}_g) \mathbf{\Omega}_g^{-1(t)} (\bxi_{ng}^{(t+1)} - \mathbf{M}^{(t+1)}_g)^{\prime}}{p n_g^{(t+1)}}\\
&~~~+\frac{\sum_{n=1}^N \hat{Z}_{ng}^{(t+1)} \bDelta_{ng}^{(t+1)}\text{tr}\left\{\bOmega^{-1(t)}\bKappa_{ng}^{(t+1)}\right\}}{p n_g^{(t+1)}},\\
\mathbf{\Omega}^{(t+1)}_{g} &= \frac{\sum_{n=1}^N \hat{Z}_{ng}^{(t+1)} (\bxi_{ng}^{(t+1)} - \mathbf{M}^{(t+1)}_g)^{\prime} \mathbf{\Phi}_g^{-1(t+1)}(\bxi_{ng}^{(t+1)} - \mathbf{M}^{(t+1)}_g)}{r n_g^{(t+1)}}\\
&~~~+\frac{\sum_{n=1}^N \hat{Z}_{ng}^{(t+1)} \bKappa_{ng}^{(t+1)}\text{tr}\left\{\bPhi^{-1(t+1)}_g\bDelta_{ng}^{(t+1)}\right\}}{r n_g^{(t+1)}}.
\end{align*}

\end{enumerate}

\subsubsection{Hybrid Approach}
While the MCMC based approach can generate exact results, fitting such models can take \underline{\textbf{21 hours}} --- on a \textbf{supercomputer} at the SciNet HPC Consortium for a dataset with $N = 1000$ and $rp = 6$ \citep{Ponce2019} --- as we need to evaluate the expected complete-data log-likelihood with respect to the posterior distribution of the latent variables at every iteration of the EM algorithm. On the other hand, the VGA approach is computationally efficient, e.g., on a single dataset from Simulation~1 with $N = 1000$ and $rp = 6$, fitting such a model takes approximately \underline{\textbf{one minute}} on a standard laptop with Apple M1~chip. However, it does not guarantee an exact posterior \citep{ghahramani1999}. Thus, we provide a computationally efficient hybrid approach in which: 
\begin{itemize}[topsep=0pt]
\item[--] Step 1: Fit the model using the VGA based approach.
\item[--] Step 2: Estimate the component indicator variable $Z_{ng}$ conditional on the parameter estimates from the VGA based approach.
\item[--] Step 3: Using the parameter estimates from Step 1 as the initial values for the parameters and using the classification from Step 2, compute the MCMC based expectation for the latent variable $\thetav_{ng}$ and obtain the final estimates of the model parameters.
\end{itemize}
The hybrid approach comes with a substantial reduction in computational overhead compared to a traditional MCMC EM but it can generate samples from the exact posterior  distribution. Fitting such a model using the hybrid approach on a single dataset from Simulation 1 with $N = 1000$ and $rp = 6$ takes on average about \underline{\textbf{6 minutes}} on a standard laptop with Apple M1 chip. When the primary goal is to detect the underlying clusters (which is the case for our the real data analysis), the VGA based approach is sufficient. However,  when the primarily goal is posterior inference, we recommend the hybrid approach as it can better yield an exact posterior similar to the MCMC-EM approach but is computationally efficient.

Details on the convergence criteria, initialization, and parallel implementation for an MCMC-EM approach is provided in Appendix~C.

\subsection{Identifiability}
Model identifiability is vital to obtain unique and consistent parameter estimates. Identifiability of univariate and multivariate finite mixtures of normal distributions has been proved \citep{teicher1963, yakowitz1968}. With regards to the mixtures of MVPLN distributions, the estimates for $\mathbf{\Phi}_{g}$ and $\mathbf{\Omega}_{g}$ are only unique up to a strictly positive constant. To eliminate identifiability issues, a constraint needs to be imposed, e.g., the trace of $\mathbf{\Omega}_{g}$ can be set equal to $p$, the trace of $\mathbf{\Phi}_{g}$ can be set equal to $r$, or the first diagonal element of $\mathbf{\Phi}_{g}$ can be set equal to~1. The latter solution, which is used by \cite{Gallaugher2017}, is used for all analyses in this paper. To obtain final parameter estimates, the resulting $\mathbf{\Phi}^{(t)}_{g}$ is divided by the first diagonal element of $\mathbf{\Phi}^{(t)}_{g}$, and $\mathbf{\Omega}^{(t)}_{g}$ is multiplied by the first diagonal element of $\mathbf{\Phi}^{(t)}_{g}$.

\subsection{Model selection and performance assessment}
Four model selection criteria are offered, which include the Akaike information criterion \citep[$\textsc{AIC}$;][]{akaike1973}, the Bayesian information criterion \citep[$\textsc{BIC}$;][]{schwarz1978}, a variation of the $\textsc{AIC}$ used by \citet{Bozdogan1994} called $\textsc{AIC3}$, and the integrated completed likelihood \citep[$\textsc{ICL}$;][]{biernacki2000}. These criteria are given by $\textsc{AIC} = -2 \log \mathcal{L} (\boldsymbol{\vartheta}^{(MLE)} |\boldsymbol{y}) + 2K$, $\textsc{BIC} = -2 \log \mathcal{L} (\boldsymbol{\vartheta}^{(MLE)} |\boldsymbol{y}) + K \log N$, $\textsc{AIC3} =  -2 \log \mathcal{L} (\boldsymbol{\vartheta}^{(MLE)}|\boldsymbol{y}) + 3K$, and $\textsc{ICL} \approx \textsc{BIC} + 2 \sum_{n=1}^N \sum_{g=1}^G \text{MAP}\{ z^{(t)}_{ng}\} \log z^{(t)}_{ng}$, respectively, where $\mathcal{L} (\boldsymbol{\vartheta}^{(MLE)} |\mathbf{y})$ represents maximized log-likelihood, $\boldsymbol{\vartheta}^{(MLE)}$ is the maximum likelihood estimate of the model parameters $\boldsymbol{\vartheta}$, $N$ is the number of observations, and $\text{MAP}\{ z^{(t)}_{ng}\}$ is the maximum $\textit{a posteriori}$ classification given $z^{(t)}_{ng}$. 

In situations where the true classes are known but, for clustering purposes, are ignored, the adjusted Rand index \citep[ARI;][]{hubert1985} can be used for performance assessment. The ARI takes a value~1 under perfect class agreement and has expected value~0 under random classification.

\begin{sidewaystable}
\centering
\caption{Number of clusters selected (average ARI, standard deviation) for each simulation setting using different model selection criteria}
\label{table_gmodelselection_sim1to3}
\scalebox{0.85}{
\begin{tabular}{lllllll}
\hline 
Method                             & Setting & $\textsc{BIC}$      & $\textsc{ICL}$      & $\textsc{AIC}$         & $\textsc{AIC3}$                & NA               \\ \hline
%   Mixtures of MVPLN   & 1       & 1 (1.00, 0.00)      & 1 (1.00, 0.00)      & 1, 3 (0.38, 0.52)      & 1, 3 (0.5, 0.53)    &                    \\
%    (MCMC based approach)                                & 2       & 2, 3 (0.88, 0.23)   & 2, 3 (0.99, 0.01)   & 2, 3 (0.66, 0.26)      & 2, 3 (0.84, 0.27)   &                    \\
%                              & 3       & 2 (1.00, 0.00)      & 2 (1.00, 0.00)      & 2 (1.00, 0.00)         & 2 (1.00, 0.00)      &                    \\ \hline
Mixtures of MVPLN                                      & 1       & 1(1.00, 0.00)       & 1(1.00, 0.00)       & 1, 2(0.88, 0.33)       & 1(1.00, 0.00)       &                    \\
(VGA based approach)                 & 2       & 2(1.00, 0.00)       & 2(1.00, 0.00)       & 2, 3, 4, 5(0.89, 0.22) & 2(1.00, 0.00)       &                    \\
    & 3       & 2(1.00, 0.00)       & 2(1.00, 0.00)       & 2, 3, 4, 5(0.86, 0.21) & 2, 3, 4(0.96, 0.12) &                    \\ \hline
\multirow{3}{*}{${\tt HTSCluster}$}& 1       & 3 (0.00, 0.00)      & 3 (0.00, 0.00)      & 3 (0.00, 0.00)         & 3 (0.00, 0.00)      &                    \\
                                   & 2       & 3 (-0.0091, 0.022)    & 3 (-0.0091, 0.022)    & 3 (-0.0091, 0.022)       & 3 (-0.0091, 0.022)    &                    \\
                                   & 3       & 3 (-0.0035, 0.0078) & 3 (-0.0035, 0.0078) & 3 (-0.0035, 0.0078)    & 3 (-0.0035, 0.0078) &                    \\ \hline
\textit{k}-means                            & 2       &                     &                     &                     &                     & 2 (-0.033, 0.024)  \\
                                   & 3       &                     &                     &                     &                     & 2 (-0.0051, 0.015) \\ \hline
                                   
\end{tabular}
}
\end{sidewaystable}

\section{Results}
\label{sec:results}
\subsection{Simulations}
Simulation studies were conducted to illustrate the ability to recover the true underlying parameters for the mixtures of MVPLN algorithm. For Simulation~1, datasets with $G = 1$ component were generated with $N = 1000$ observations, $r = 2$ and $p = 3$. For Simulation~2, datasets with $G = 2$ components and $\pi_1 = 0.79$ were generated with $N = 1000$ observations, $r = 2$ and $p = 3$. For Simulation~3, datasets with $G = 2$ components and $\pi_1 = 0.6$ were generated with $N = 1000$ observations, $r = 2$ and $p = 3$. Further, only diagonal covariance structures for both $\mathbf{\Phi}_g$ and $\mathbf{\Omega}_g$ were considered in Simulation~3. Each of the simulation settings consisted of $25$ different datasets. The count range in the simulated datasets closely represented the range observed in the RNA-seq data \citep{freixascoutin2017}. The covariance matrices $\mathbf{\Phi}_g$ and $\mathbf{\Omega}_g$ for each setting are generated using the ${\tt clusterGeneration}$ package \citep{clusterGeneration2015}. Initialization of $z_{ng}$ was done using one hundred different runs of the $k$-means algorithm. Clustering was performed on each dataset for values $G = 1,2, 3$.

Comparative studies were also conducted. Because no comparable methods capable of clustering three-way count data were found in the current literature, datasets from Simulations~1, 2 and 3 were vectorized and analyzed with clustering methods designed for two-way data. For this purpose, a model-based clustering technique for count data, {\tt HTSCluster} \citep{rau2011, rau2015}, and a distance-based method, \textit{k}-means clustering \citep{MacQueen1967}, were used. For {\tt HTSCluster}, initialization and clustering ranges were same as those used for mixtures of MVPLN algorithm. For \textit{k}-means, the algorithm was run by specifying the true number of clusters and only the partitioning of the observations into clusters was evaluated. For this reason \textit{k}-means was not performed for Simulation~1. 

The clustering results along with ARI values of our proposed method and other comparative methods are provided in Table~\ref{table_gmodelselection_sim1to3}. In all three Simulations, our proposed approach was able to recover the underlying cluster in all 25 datasets using both BIC and ICL. Note that we only provide the ARI values from the VGA approach. The ARI from the hybrid approach is the same as that from the VGA approach because the cluster membership indicator variable is determined in the VGA step in the hybrid approach. We do not provide the fitted values for the MCMC-EM due to the extreme computational cost that comes with fitting these models. The parameter estimation results for $\boldsymbol{M}_g$, $\mathbf{\Phi}_g$ and $\mathbf{\Omega}_g$ via the mixtures of MVPLN algorithm for Simulations~1, 2 and~3 using both the VGA approach and the hybrid approach are summarized in Appendix~A. As can be seen through the simulations, both the VGA and hybrid approach can recover the parameter estimates very well. However, there is a slight increase in the precision of these estimations with the hybrid approach.
Overall, the simulation experiments illustrated that our approach for parameter estimation (Section~\ref{sec:para}) is effective at parameter recovery for the mixtures of MVPLN distributions. With regards to {\tt HTSCluster}, a model with $G = 3$, the highest value for $G$ considered, was selected for all simulation settings. Furthermore, the ARI values were low across all simulation settings, indicating that observations were not assigned to the correct clusters. For \textit{k}-means clustering, the ARI values were also low despite the algorithm being given the correct value for $k$ \textit{a~priori}.

\subsection{Clustering transcriptome data} 
To illustrate the applicability of mixtures of MVPLN distributions for detecting the underlying cluster structure, the VGA based approach was applied to a RNA-seq dataset. Typically, only a subset of genes from the experiment are used for cluster analysis, in order to reduce noise. For this analysis, only the differentially expressed genes were used for clustering. \citet{freixascoutin2017} used RNA-seq to monitor the transcriptional dynamics in the seed coats of darkening and non-darkening cranberry beans (\textit{P. vulgaris}) at three developmental stages: early, intermediate and mature. The aim of the study was to evaluate if the changes in the seed coat transcriptome were associated with proanthocyanidin levels as a function of seed development in cranberry beans. The RNA-seq data are available on the National Center for Biotechnology Information (NCBI) Sequence Read Archive (SRA) under the BioProject PRJNA380220.

\begin{figure}[!h]%figure3
\centerline{\includegraphics[width=0.45\textwidth]{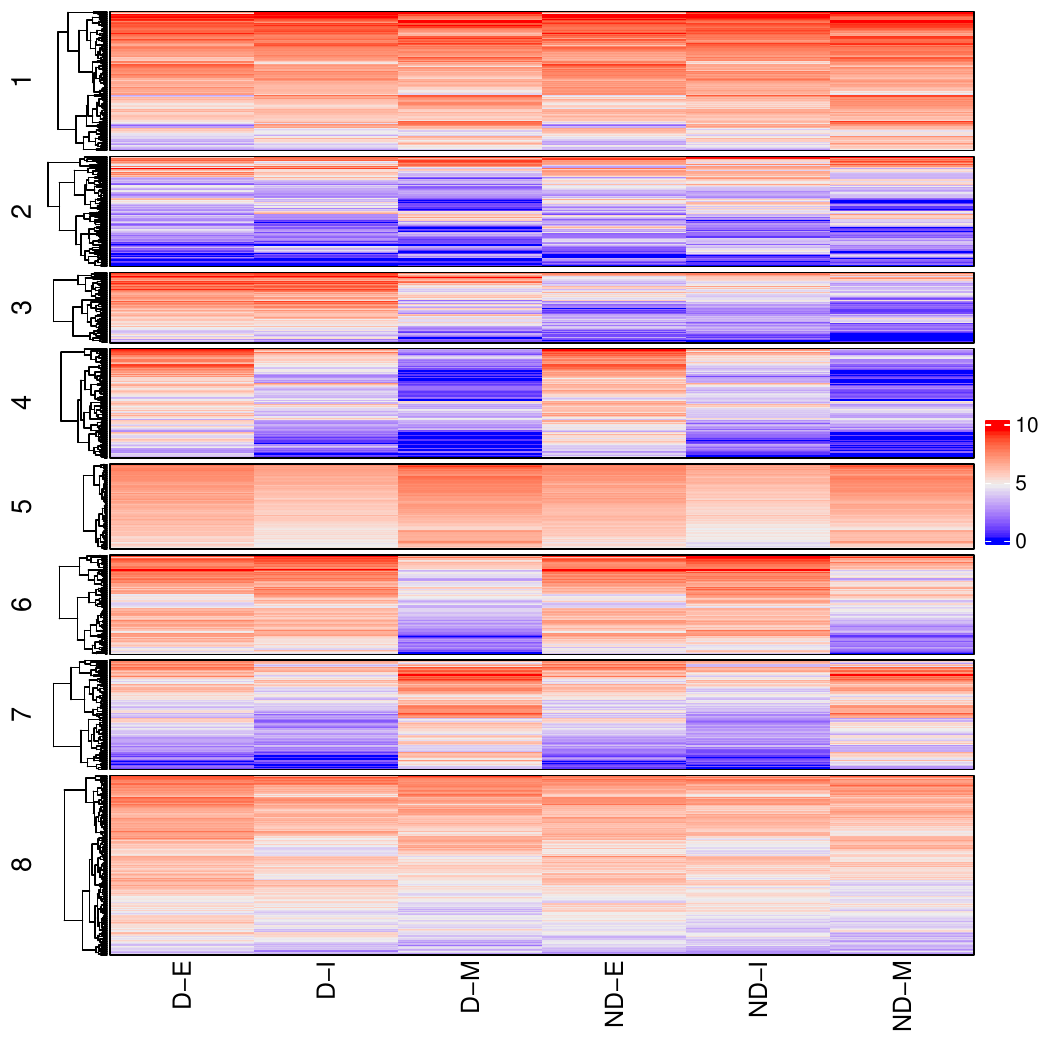}}
\caption{Heatmap showing, for Cluster~1 through Cluster~8, log-transformed gene expression patterns for the $G = 8$ model selected by both BIC and ICL for the cranberry bean RNA-seq dataset. The red and blue colors represent the log-transformed expression levels, where red represents high expression and green represents low expression. The rows represent the genes and the columns represent samples involved in the RNA-seq study. Respectively, the samples are D-E: darkening early, D-I: darkening intermediate, D-M: darkening mature, ND-E: non-darkening early, ND-I: non-darkening intermediate, and ND-M: non-darkening mature cranberry bean.}\label{MVPLN_bozzo_heatmap_cluster}
\end{figure}

The study identified $1336$ differentially expressed genes, which were used for clustering. The raw read counts for genes were obtained from Binary Alignment/Map files using samtools \citep{li2009} and HTSeq \citep{anders2015}. The median value from the $3$ replicates per each developmental stage was used. On the three-way data of dimensions $1336 \times 2 \times 3$, a clustering range of $G = 1, \ldots,10$ was considered using $k$-means initialization (100 runs). Furthermore, we repeated the analysis ten times. Since BIC and ICL both performed well in recovering the underlying cluster structure in the simulated data, here we also used BIC and ICL for model selection. Both BIC and ICL selected a model with $G = 8$. In this model, Clusters~1-8 were composed of $206$ $(15.4\%)$, $163$ $(12.2\%)$, $104$ $(7.8\%)$, $162$ $(12.1\%)$, $126$ $(9.4)$, $147$ $(11.0)$, $162$ $(12.1)$, and $266$ $(19.9)$ genes, respectively.  See Appendix~B for gene composition of each cluster. Expression patterns across the clusters were visualized using a heatmap. The log-transformed expression patterns of the clusters are illustrated using the heatmap in Figure~\ref{MVPLN_bozzo_heatmap_cluster}.  

\begin{figure}[!h]%figure3
\centerline{\includegraphics[width=0.45\textwidth]{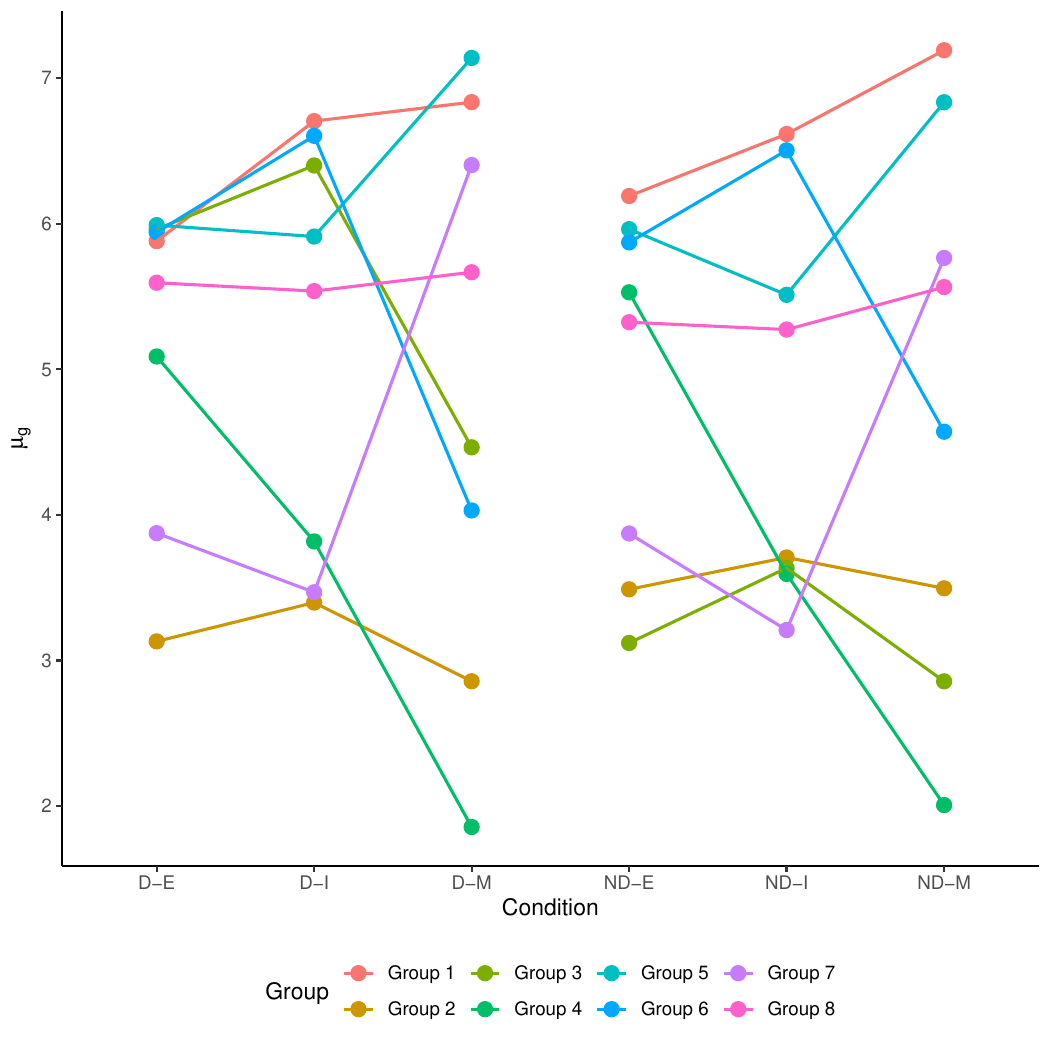}}
\caption{Visualization of the cluster-specific  $\mathbf{\mu}_g$ which relates to the mean trends of the expression levels of genes in Cluster~1 to Cluster~8.
}\label{mean_trends}
\end{figure}

Because no true labels are available for assessing the transcriptome data cluster analysis results, the clustering results were explored using the cluster-specific $\mathbf{\mu}_g$ which relates to mean trends of the expression levels of genes in different clusters in Figure~\ref{mean_trends}. As evident in Figure~\ref{mean_trends}, each cluster has its distinctive expression signatures. In some clusters (for example, Cluster~4), the means of the gene expression signatures are similar between the darkening and non-darkening beans but their mean expression pattern varies between developmental stages; whereas, in Cluster~8, the means of the gene expression signatures are similar across the developmental stages but varies slightly between the darkening and non-darkening beans. On the other hand, in Cluster~3, the mean gene expression signatures vary across both the developmental stages as well as between darkening and non-darkening beans.

For all simulation and transcriptome data analyses, the normalization factors representing library sizes for samples were obtained using the trimmed mean of \textit{M} values from {\tt calcNormFactors} function of {\tt edgeR} package \citep{robinson2010, mccarthy2012}. 

A table of mathematical notation is provided in Appendix~D and a table of abbreviations is provided in Appendix~E. 

\section{Discussion}
A mixture of MVPLN distributions is introduced for clustering three-way count data, targeted at expression data arising from RNA-seq experiments. This is the first use of a mixture of MVPLN distributions for clustering within the literature. By allowing for a direct analysis of three-way data structures, matrix variate distributions permit the estimation of correlations within and between variables and occasions. This makes them very attractive for analyzing matrix data in the context of clustering. Further, by considering a matrix variate structure, the number of free covariance parameters to be estimated is greatly reduced under high dimensional settings. Herein, three different parameter estimation frameworks are proposed: an approach based on MCMC, on based on VGA, and a hybrid approach. When posterior inference is of interest, the MCMC based approach is favourable but it can be computationally intensive. On the other hand, the VGA based approach only approximates the posterior distribution that relies on approximation but it is computationally efficient. Therefore, here we also propose a hybrid approach that is computationally efficient and it samples from the true posterior. Through simulation studies, we show that the VGA approach provides good clustering performance.

Using simulated data, it was illustrated that the algorithm for mixtures of MVPLN distributions is effective and returned favorable clustering results. The transcriptome data analysis showed the applicability of the mixture model-based clustering method on RNA-seq count data. A possible future direction of this work would be to make use of subspace clustering methods and to develop the matrix variate factor analyzers model. This would permit clustering of data in low-dimensional subspaces as high-dimensional RNA-seq datasets become frequent. Another path is to consider restrictions on the matrices $\mathbf{\Phi}_g$ and $\mathbf{\Omega}_g$, as done by \citet{Viroli2011}. Also, constraints on $\mathbf{\Phi}_g$ similar to those introduced by \cite{McNicholas2010a}, and used by \citet{Anderlucci2015}, could be beneficial when analyzing longitudinal RNA-seq data.

\section*{Acknowledgements}
The authors acknowledge the computational support by Dr. Marcelo Ponce at the SciNet HPC Consortium, University of Toronto, ON, Canada.
\section*{Funding}
This work was supported by Discovery Grant from the Natural Sciences and Engineering Research Council of Canada (NSERC) [to SS], Canada Research Chairs program [to SS and PM], Queen Elizabeth II Graduate Scholarships in Science \& Technology [to AS], Ontario Graduate Fellowship [to AS], and an E.W.R. Steacie Memorial Fellowship from NSERC [to PM]. No funding body played a role in the design of the study, analysis and interpretation of data, or in writing the manuscript. \\ \\
\noindent \textbf{Conflict of Interest}: None declared.

%\bibliographystyle{agsm}
%\bibliography{bibofA}

\includepdf[pages=-]{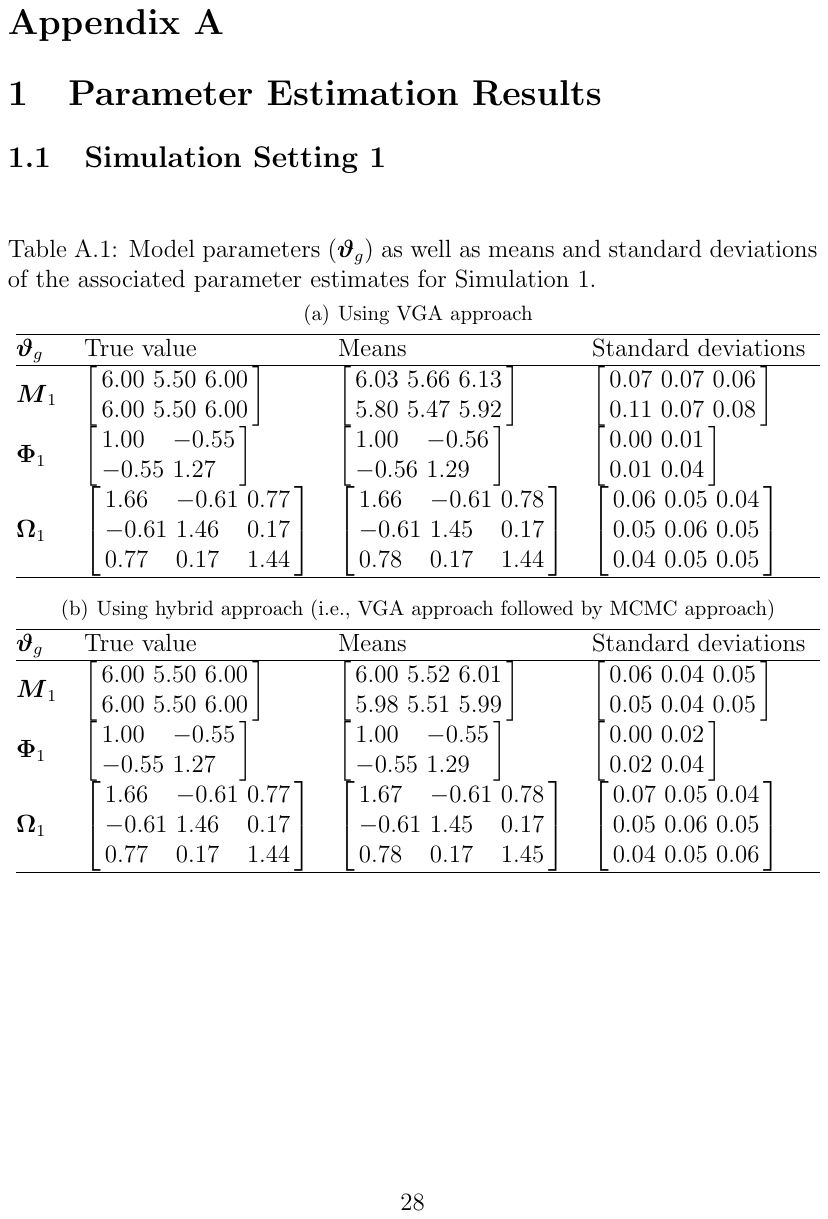}
\includepdf[pages=-]{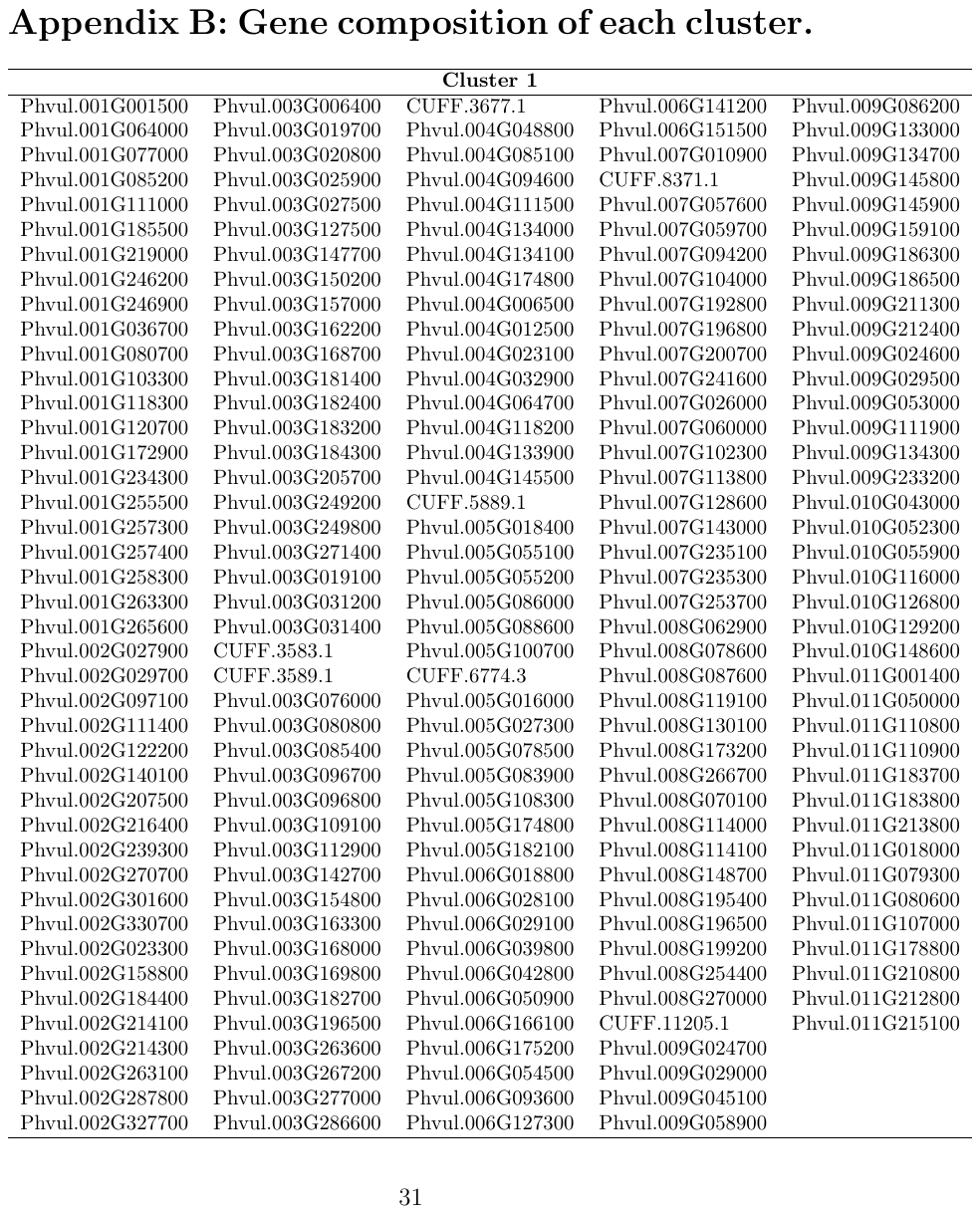}
\includepdf[pages=-]{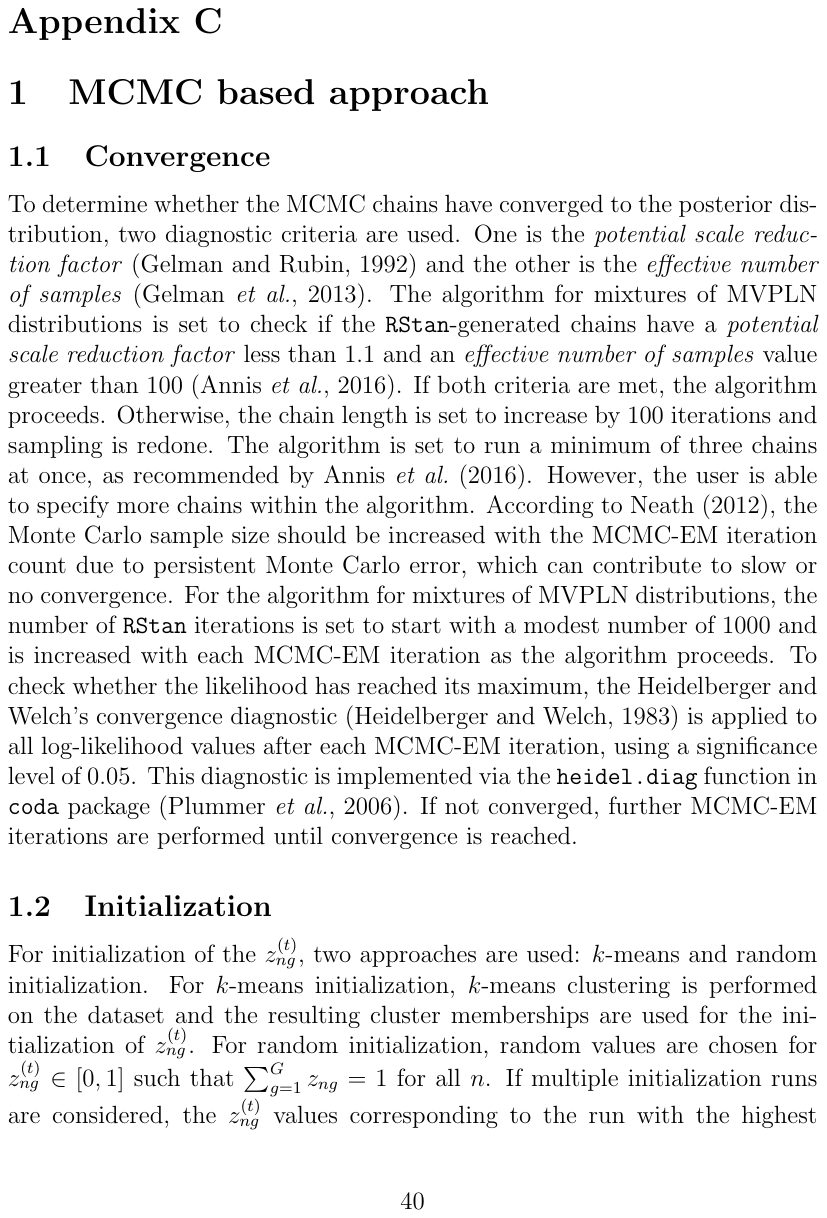}
\includepdf[pages=-]{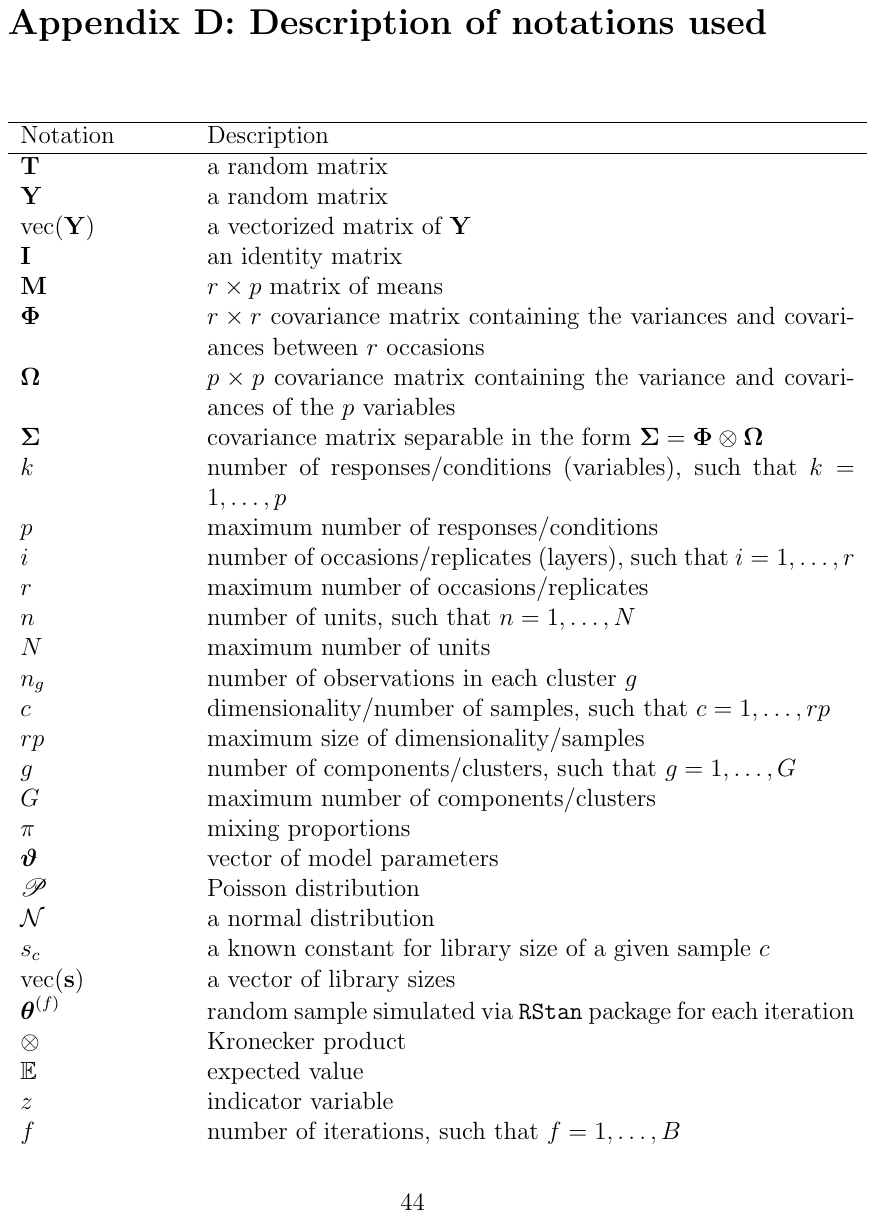}
\includepdf[pages=-]{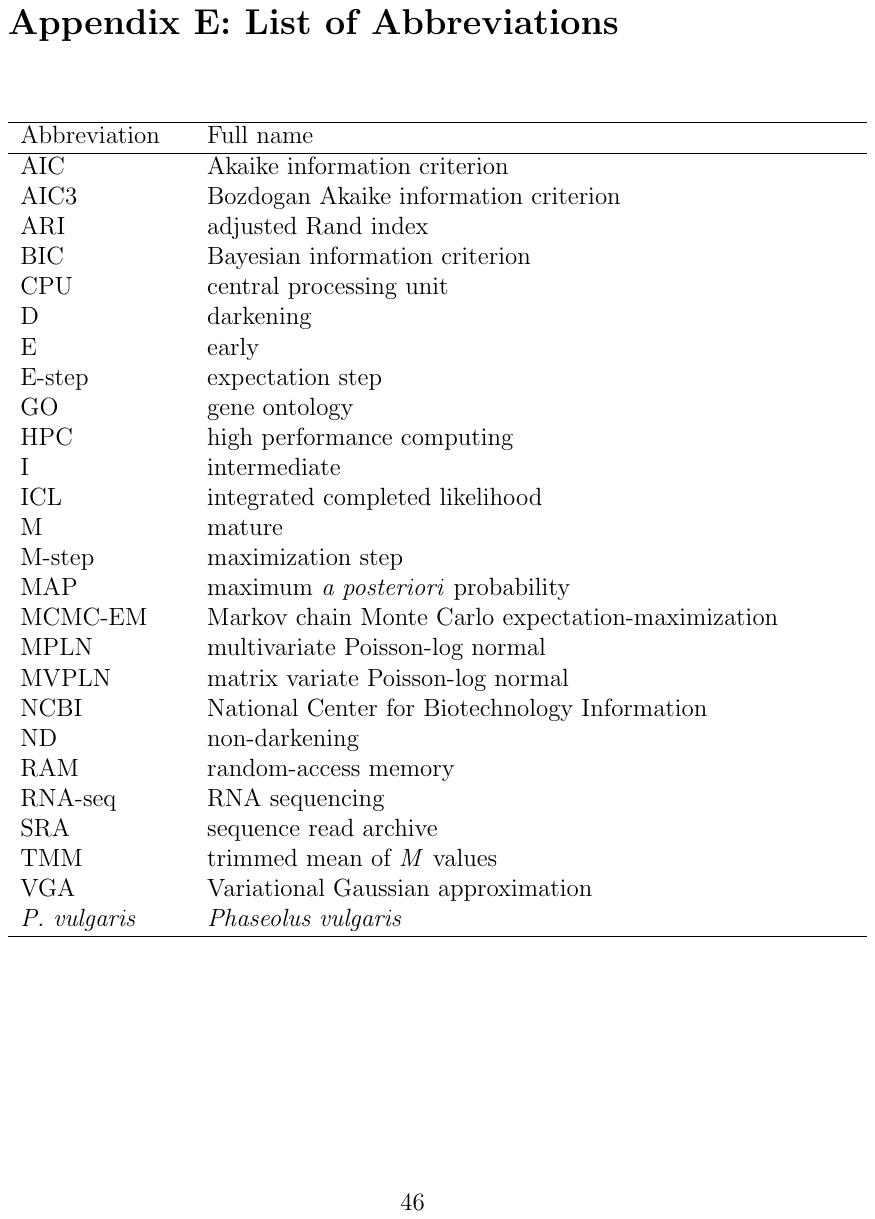}

\end{document}